\documentclass[twocolumn]{article}
\usepackage[compress]{cite}
\usepackage{IJCIR,multicol,times}

\usepackage{epsfig}

\begin{document}
\pagestyle{empty}
\sloppy

\title{A Fault Tolerant Trajectory Clustering (FTTC) for selecting cluster heads in Wireless Sensor Networks}

\author{{\bf Hazarath Munaga$^1$, J. V. R. Murthy$^1$ and N. B. Venkateswarlu$^2$} \\[1em]
$^1$Department of Computer Science and Engineering, UCEK\\
J.N.T.U Kakinada, Andhra Pradesh, India\\
\textit{\{hazarath.munaga,mjonnalagedda\}@gmail.com} \\[1em]
$^2$Department of Computer Science and Engineering,\\
AITAM, Tekkali, Andhra Pradesh, India\\
\textit{venkat\_ritch@yahoo.com} \\[1em]
}

\maketitle

\begin{abstract}
Wireless sensor networks (WSNs) suffers from the hot spot problem where the sensor nodes closest to the base station are  need to relay more packet than the nodes farther away from the base station. Thus, lifetime of sensory network depends on these closest nodes.  Clustering methods are used to extend the lifetime of a wireless sensor network. However, current clustering algorithms usually utilize two techniques; selecting cluster heads with more residual energy, and rotating cluster heads periodically to distribute the energy consumption among nodes in each cluster and lengthen the network lifetime. Most of the algorithms use random selection for selecting the cluster heads. Here, we propose a Fault Tolerant Trajectory Clustering (FTTC) technique for selecting the cluster heads in WSNs. Our algorithm selects the cluster heads based on traffic and rotates periodically. It provides the first Fault Tolerant Trajectory based clustering technique for selecting the cluster heads and to extenuate the hot spot problem by prolonging the network lifetime.
\end{abstract}
\\\\

\section{Introduction}

Wireless sensor networks ({\it hereinafter}, WSNs) are networks of wireless nodes that are deployed over an area for the purpose of monitoring certain phenomena of interest. To keep specific areas under observation, WSNs deploy hundreds or thousands of integrated sensor nodes to sample data from observed environment. Although, these devices are not very accurate and reliable individually, their deployment in large number enhances their accuracy and reliability. In addition WSNs can provide area coverage that was not possible with other wireless networks. They can be deployed in extremely hostile environments, such as near volcanically active sites, inside a chemical industry, or probable disaster areas or in mainly inaccessible environments. \\\\
\par WSNs have numerous advantages. They are easier and faster to deploy than any wired network. They have a large coverage area and longer range. They have higher degree of fault tolerance than wireless networks, because failure of one or more sensors or nodes does not effect the operation of the network and mainly they are self-configuring. The nodes perform certain measurements, process the measured data and transmit the processed data to a base station over a wireless channel. The base station collects data from all the nodes, and analyzes this data to draw conclusions about the activity in the area of interest. In practice, due to the large quantity of sensor nodes, it is infeasible to recharge the batteries in WSNs. Therefore, sensor network lifetime is a primary concern in sensor network design.\\
\par In literature many researchers proposed various protocols to reduce the energy consumption and improve the network lifetime of WSNs. Those protocols can be categorized into three classes: routing protocols, sleep-and-awake scheduling protocols, and clustering protocols. The routing protocols \cite{1}
\cite{2}\cite{3} 
\cite{4} determine the energy-efficient multi-hop paths from each node to the base station. In sleep-and-awake scheduling protocols 
\cite{5}
\cite{6}\cite{7}, every node in the schedule can sleep, in order to minimize energy consumption. In clustering protocols \cite{8}\cite{9}\cite{10}\cite{11}
\cite{12}\cite{13}
\cite{14}
\cite{16},
 data aggregation can be used for reducing energy consumption. Data aggregation, also known as data fusion, can combine multiple data packets received from different sensor nodes. It reduces the size of the data packet by eliminating the redundancy \cite{17}
\cite{18}
. Wireless communication cost is also decreased by the reduction in the data packets. Hence, by reducing the energy consumption clustering protocols increases the network lifetime.\\
\par Clustering \cite{19} is a commonly adopted approach in sensor networks to manage power efficiently. In clustering, sensors in the monitoring area are grouped into clusters; all sensor nodes within the same cluster send their data to the cluster head, which then forwards the aggregated data to the base station. In this way clustering reduces the overhead and increases the network lifetime; on the other hand, the disadvantage is due to heavy usage of cluster heads ``typically cluster heads die at an early stage'' \cite{20}. This is sometimes called the hot spot problem \cite{21}. Without adding extra nodes or redistributing the available energy, this problem is hard to solve. For example, Ref. \cite{20} has shown that, with varying transmission power of nodes and even considering unlimited transmission ranges does not solve the hot spot problem. At the same time, it is also envisioned that sensor nodes will become ``extremely inexpensive'' \cite{22}. While beyond a certain node density, adding additional nodes does not provide any improvement regarding sensing, communication or coverage \cite{23}, adding nodes might obviously help to increase the lifetime of a sensor network while providing the same service to its users, i.e. leveraging sensor values from the same number of nodes. \\
\par Ref. \cite{14} proposed the randomized clustering algorithm to organize sensors into clusters in a wireless sensor network. Computation of the optimal probability of becoming a cluster head was presented. Ref. \cite{24} defined the maximum cluster-lifetime problem, and they proposed distributed, randomized algorithms that approximate the optimal solution to maximize the lifetime of dominating sets on wireless sensor networks., \cite{25} considered the k-domatic partition problem, and they proposed three deterministic, distributed algorithms for finding large k-domatic partitions.\\
\par Ref. \cite{8} proposed LEACH, a well-known clustering protocol for wireless sensor networks. LEACH includes distributed cluster formation, local processing to reduce global communication and randomized rotation of cluster heads among all the nodes in the network. Each cluster selects a cluster head, which is responsible for aggregating collected data and sending data to base station. LEACH provides a good model that helped to reduce information overloaded and provides a reliable data to the end user. Together, these features allow LEACH to achieve the desired properties. Also, an improved scheme of LEACH was proposed, named LEACH-C \cite{9}. In LEACH-C, a centralized algorithm at the base station makes cluster formation.\\

\par Ref. \cite{26} deals with the problem of finding an energy-balanced solution to data propagation in WSNs using a probabilistic algorithm was considered for the first time. The lifespan of the network is maximized by ensuring that the energy consumption in each slice is the same. Sensors are assumed to be randomly distributed with uniform distribution in a circular region or, more generally, the sector of a disk. Data have to be propagated by the WSN towards a sink located at the centre of the disk, and it is shown that energy balance can be achieved if a recurrence relation between the probabilities that a slice ejects a message to the sink is satisfied.\\
\par Ref. \cite{10} proposed PEGASIS. PEGASIS makes a communication chain using a Traveling Sales Person heuristic. In PEGASIS, nodes are organized into a chain using a greedy algorithm so that each node transmits to and receives from one of its neighbors. A randomly selected node from the chain will forward the aggregated data to the base station, thereby reducing per round energy expenditure compared to LEACH.\\
\par Ref. \cite{27} proposed clustering-based routing protocol called base station controlled dynamic clustering protocol (BCDCP), which utilizes a high energy base station to set up cluster heads and perform other energy-intensive tasks, can noticeably enhance the lifetime of a network. \\
\par Ref. \cite{28} proposed two new algorithms under the name of PEDAP, which are near optimal minimum spanning tree based wireless routing schemes. The performance of the PEDAP was compared with LEACH and PEGASIS, and showed a slightly better network lifetime than PEGASIS. Ref. \cite{29} proposed a new routing scheme; called SHORT, to achieve higher energy efficiency, network lifetime, and more throughput than PEGASIS, and PEDAP-PA protocols. This scheme used the centralized algorithms and required the powerful base station. The performance results showed that SHORT can achieve better ``energy X delay'' performance than the existing chain based data aggregation protocols. \\
\par Ref. \cite{11} proposed HEED, by extending LEACH and considering range limits of the wireless communication and intra-cluster communication cost. The probability for each node to become a tentative cluster head depends on its residual energy, and all the tentative heads in which are competing for becoming the final cluster heads. The final cluster heads are selected according to the intra-cluster communication cost. HEED terminates within a constant number of iterations, and achieves fairly uniform distribution of cluster heads across the network.\\
\par Ref. \cite{30} proposed EECR, which is an energy efficient clustering routing algorithm. The performance of the EECR was compared with LEACH, and showed a slightly better network lifetime than LEACH.\\
\par However, the unsolved problem of considerable energy consumption on the cluster formation still exists. Here, we consider the path followed by the node/sensor to transfer data to the base station as the ``{\it trajectory}'', and using the proposed trajectory clustering \cite{31}\cite{32}, we cluster the trajectories and obtain the ``{\it representative trajectory}''. Then the  nodes lies in the representative trajectory are considered as the cluster heads and the obtained cluster heads will be used for communicating data to the base station. Moreover, we concentrated on the rotation of cluster heads among all sensor nodes to improve the lifetime of the network based on the traffic density. We tested our proposed method and found that this method considerably enhances the lifetime of the network.
\section{Novel Algorithm}
This section considers the wireless sensor networks consisting of hundreds or thousands of deployed sensor nodes in the sensing field. On the basis of \cite{10}\cite{27}, it is assumed by the following properties of the wireless sensor networks to simplify the network model. 
\begin{itemize}
\item The base station is located far away from the sensors,
\item The nodes have uniform initial energy allocation and all sensor nodes have equal capabilities (data processing, wireless communication, battery power).
\item All sensor nodes have various transmission power levels, and each node can change the power level dynamically. 
\item Each node senses the environment at a fixed rate, and
\item All nodes are immobile.
\end{itemize}
\par The sensor nodes are geographically grouped into clusters and capable of operating in two basic modes: the sensing mode and the cluster head mode \cite{10}. In the sensing mode, the node senses the task and sends the sensed data to its cluster head. In cluster head mode, a node gathers data from its cluster members, performs data fusion, and transmits the data to the base station. The base station in turn performs the key task of cluster head selection. 
\subsection{Cluster head selection}
Initially the nodes will transmit a hello packet to the base station. For calculating the shortest path, we can consider various Quality of service (QoS) parameters like spatial distance, processing delay, available bandwidth, allocated buffer space etc., in this phase of research we consider only spatial distance. Then, we used Dijkstra's \cite{33} algorithm for finding the shortest path of the hello packet. After receiving hello packets from the nodes, using the Trajectory Clustering algorithm, the base station computes the representative trajectory by clustering the trajectories (here the trajectory is nothing but the path used by the node to transfer its data to the base station). The nodes of the obtained representative trajectory are considered as the cluster heads. Then the base station splits the network into clusters (equal to the number of nodes in the representative trajectory), and identifies the nodes in the representative trajectory as the corresponding cluster heads. Then, the base station broadcast a message to the network mentioning about the nodes and their corresponding cluster heads. Subsequently the nodes will use its cluster heads to transmit its data. This process will be performed periodically and the cluster heads will change based on the traffic. \\
\par Cluster head selection routine contains the following stages:-
\begin{enumerate}
\item Base station computes the cluster heads using TC algorithm;
\item Split the network into N clusters; and
\item Broadcast message to all nodes mentioning cluster members and their corresponding cluster heads
\end{enumerate}
\subsection{Trajectory Clustering}
The success of any clustering algorithm depends on the adopted dissimilarity measure. Following section explains about the adopted dissimilarity measure.\\
\par 
Agrawal et al., \cite{34}, proposed the usage of {\it Euclidean distance} between time series of equal length as the measure of their similarity. The idea has been generalized in \cite{35} for subsequence matching. In a similar way \cite{36} used {\it Discrete Wavelet Transform} and \cite{37} used {\it Principal Component Analysis} for measuring time series similarity. Another approach which is brought from image processing is {\it Time Warping technique} and it is used in \cite{38} to match signals in speech recognition. A similar technique is used to find longest common subsequence (LCSS) of two sequences using fast probabilistic algorithms to compute the LCSS, and then define the distance using the length of this subsequence \cite{39}. In \cite{40} suggested this technique to measure the similarity of time-series data in data mining. \\
\par
Here we adopted {\it Hausdorff distance} \cite{41} for calculating dissimilarity between trajectories. The following are some of the definitions used in our algorithm.\\
\par
{\it Definition 1}: A {\it trajectory (t)} is represented as trj({\(t_{id}\)},{\(u_0\)},{\(u_1\)},{\(u_2\)}..,{\(u_n\)}) where ({\(t_{id}\)}) is a unique trajectory id (data packet), and ({\(u_0\)},{\(u_1\)},{\(u_2\)},..,{\(u_n\)}) is a sequence of nodes reflecting the spatial position of the node.\\
\par
{\it Definition 2}: We define the {\it spatial dissimilarity} function between two trajectories {\(t_1\)} and {\(t_2\)} as the maximum of one way distances between two trajectories. \\
The one way distance from a trajectory {\(t_1\)} to another trajectory {\(t_2\)} is defined as the integral of the Hausdorff distance between points of {\(t_1\)} to trajectory {\(t_2\)} divided by the number of points in {\(t_1\)} ({\(\mid\)}{\(t_1\)}{\(\mid\)}).
\[
{\rm dist_{ow}(t_1,t_2)  =  }\frac{{\rm 1}}{{{\rm |t_1|}}}\int\limits_{p \in t_1} {d_h} (p,t_2)dp
\]
The Hausdorff distance from a trajectory point {\it p} to another trajectory {\(t_2\)} is defined as 
$
d(p,t_2 ) = \min _{q \in t_2 } \left\{ {d(p,q)} \right\}
$.The distance between trajectories {\(t_1\)} and {\(t_2\)} is the maximum of their one way distances. 
\[
{\rm dist(t_1,t_2)  =  max\{ dist_{ow}(t_1,t_2), dist_{ow}(t_2,t_1)\} }
\]
Clearly the {\(dist_{ow}(t_1,t_2)\)} is not symmetric but {\(dist(t_1,t_2)\)} is symmetric. Note that {\(dist_{ow}(t_1,t_2)\)} is the integral of the shortest distances from points in {\(t_1\)} and {\(t_2\)}.

\subsubsection{Fault Tolerant Trajectory Cluster Routine}
Trajectories are grouped into clusters using the {\it threshold}. Here the threshold is considered as a maximum value, such that all trajectories are grouped into a single cluster. The trajectory cluster routine contains the following stages: 
\begin{enumerate}
\item Dissimilarity matrix for trajectories will be computed using the Hausdorff distance,
\item Using following {\it Initialization} Algorithm trajectories are grouped into initial clusters;
\begin{enumerate}
\item Take first sample as first cluster. Classify all the remaining trajectories into this cluster if they are within the threshold.
\item Take a trajectory (sequentially) which is not already classified into any of the cluster and consider it as a new cluster. Take all the other trajectories which are not kept in any of the clusters and keep in this cluster if they satisfy the threshold limit.
\item Repeat step b till no new clusters are added.
\end{enumerate}
\item Using the following {\it RepTraj} Algorithm representative trajectories are computed. 
\begin{enumerate}
\item For each Trajectory of cluster C calculate cumulative dissimilarity with all other trajectories of the same cluster C. Select the trajectory which is having minimum cumulative dissimilarity and take this as representative trajectory of that cluster.
\end{enumerate}
\item By considering the trajectories received from step 3, as initial cluster centers, using the follwing {\it Re-cluster} Algorithm re compute clusters and their representative trajectories until there is no change in the representative trajectories.
\begin{enumerate}
\item For each Trajectory calculate dissimilarity with all the {\it K} representative trajectories and classify to the cluster for which dissimilarity is low.
\item Re-calculate representative trajectories using {\it RepTraj} Algorithm.
\end{enumerate}
\end{enumerate}
\subsection{Fault Tolerant Routine}
\par To make the system fault tolerant, instead of selecting only one optimal representative trajectory, routine is asked to select next {\it p} number of optimal trajectories, based on their priority levels, and there corresponding cluster heads will be selected for communicating data to the base station. This list will keep by the base station, in any case of fault occurs with the initial cluster heads, as a choice it can go for other alternatives.  
\subsection{Data communication phase}
There are three steps during the data communication phase: data collection, data fusion and data transmission. Initially each sensor node transmits the sensed information to its cluster head at the time slot assigned by its cluster head. In order to save its energy, the node will close transmit part during the time slot which is not required to it. Once data from all sensor nodes have been received, the cluster head performs data fusion on the collected data and reduces the amount of raw data that need to send to the base station. Once the data gathering and data fusion are completed, the cluster head sends the compressed data to the base station.\\
\par As mentioned previously, all the nodes can work as cluster heads. Due to this, any node can become a cluster head or a cluster member. At each turn, the cluster head calculates available power and compares with the cluster members. Whenever the cluster heads power becomes less than the minimum power holding, then the cluster heads inform to its cluster members and assigns the maximum power holding cluster member as the cluster head and the same inform to the base station. Whenever the cluster head is changed, base station repeats the cluster finding process and modifies the clusters.
\section{Experimental Work}
To evaluate the performance of the algorithm, it has been simulated and compared its performance with energy efficient clustering routing (hereafter, EECR). Before the simulation and results are introduced, the radio model and some important parameters \cite{30} used in simulation have been described.
\subsection{The radio model}
We have used both the free-space propagation model and the two-ray ground propagation model to approximate path loss sustained because of wireless channel transmission. Given a threshold transmission distance of d0, the free-space model is used when ${\rm d < d_0}$ and the two-ray model is applied for cases when ${\rm d} \ge {\rm d_0}$. Using these two models, the transmit energy costs for the transfer of a b-bit data message between two nodes separated by a distance of d meters is given:\\
\[if {\rm d < d_0}, E_T \left( {b,d} \right) = E_{Tx} b + E_{amp} (d)b = E_{Tx} b + \varepsilon _1 d^2 b -(1)\]
\[if {\rm d} \ge {\rm d_0}, E_T \left( {b,d} \right) = bE_{BF} b + \varepsilon _2 d^4 b- - -(2)\]
With regard to the energy cost incurred in the receiver of the destination node, we give in Eq. (3):
\[E_T \left( b \right) = E_{Rx} b- - - (3)\]
We have summarized the different meanings and values for energy terms in Table 1. Energy consumed during data aggregation in the cluster head {\(E_{da}\)}, is also taken into account.
\begin{table*}[tb]
\renewcommand{\arraystretch}{2.0}
\caption{Summarizes meaning of each term and typical value}
\label{det}
\begin{center}
\begin{tabular}{|c|c|c|r|}
\hline S.No & Term & Meaning & Value\\
\hline\hline
1 & {\(E_{da}\)} & Consume energy for data aggregation & 5nJ/bit\\[-2ex]
2 & {\(E_{Tx}\)}, {\(E_{Rx}\)} & Radio Electronics Energy & 50nJ/bit\\[-2ex]
3 & {\(\epsilon_1\)} & Transmit applied for free space & 10pJ/(bit*$m^2$)\\[-2ex]
4 & {\(\epsilon_2\)} & Transmit applied for two way model & 0.0013 pJ/(bit*$m^4$)\\
\hline
\end{tabular}
\end{center}
\end{table*}
\subsection{The number of clusters}
We assume that N nodes are distributed in the area of A*A randomly. If there are M clusters, then there are N/M nodes in each cluster on an average. Every cluster head receives the sensed data from its cluster nodes, aggregates all the data, and sends it to the base station. \\
The total energy spent on transmitting a frame for every cluster head can be expressed as:
\[
E_1  = bE_{Tx} \frac{N}{M} + bE_{da} \frac{N}{M} + b\varepsilon _2 d_1^4 - - - (4) 
\]
where {\(d_1\)} is the distance between cluster head and base station.\\
In one frame, the cluster nodes transmit the sensed data messages to its cluster head. The energy spent for each cluster member is as below:
\[
E_2  = bE_{Tx}  + b\varepsilon _1 d_2^2 - - - (5) 
\]
where {\(d_2\)} is the distance between the member node and its cluster head. If the cluster head is in the center of the cluster, the density of every cluster is $ \rho {\rm   =  M /A2 }$, then {\(d_2\)} can equate to 
\[
d_2  = \sqrt {\frac{1}{{2\pi }}\frac{{A^2 }}{M}} - - - (6) 
\]
The energy spent for each cluster member is modified as:
\[
E_2  = bE_{Tx}  + b\varepsilon _1 \frac{1}{{2\pi }}\frac{{A^2 }}{M} - - -  (7)
\]
The energy dissipation in a cluster can be expressed as:
\[
E_2  = bE_{Tx}  + b\varepsilon _1 \frac{1}{{2\pi }}\frac{{A^2 }}{M} - - - (8)
\]
The total energies dissipated in all the clusters can be expressed as: 
\[
E = ME_c  = b\left[ \begin{array}{l}
 2E_{Tx} N + E_{da} N + M\varepsilon _2 d_1^4  \\ 
  + \left( {N - M} \right)\varepsilon _1 \frac{1}{{2\pi }}\frac{{A^2 }}{M} \\ 
 \end{array} \right] - - -(9)
\]

if, $\frac{{\partial E}}{{\partial M}} = 0$, we can get the following Eq. (10)
\[
M = A\sqrt {\frac{N}{{2\pi }}\frac{{\varepsilon _1 }}{{\left( {\varepsilon _2 d_1^4  - E_{Tx} } \right)}}}- - - (10) 
\]
In our simulation, we consider N = 100, A = 100 $m^2$ and {\(d_1\)} = 90m, and for various number of clusters i.e., from six to twelve.
\subsubsection{Results}
It has been simulated that 100 nodes randomly located in the sensing field of 100 X 100 $m^2$ with the base station located at least 90 m away. All sensor nodes periodically sense events and transmit the data packet to the base station. All sensor nodes start with an initial energy of 2 J and the data message size is fixed at 516 bytes, of which 16 bytes represent the weight value. We choose three different coefficients {\(C_1\)} = 0.5, {\(C_2\)} = 0.4 and {\(C_3\)} = 0.1. To evaluate the performance of our algorithm, we compare its performance with EECR. \\
\par
\begin{figure}[tb]
\begin{center}
\includegraphics[width=8cm]{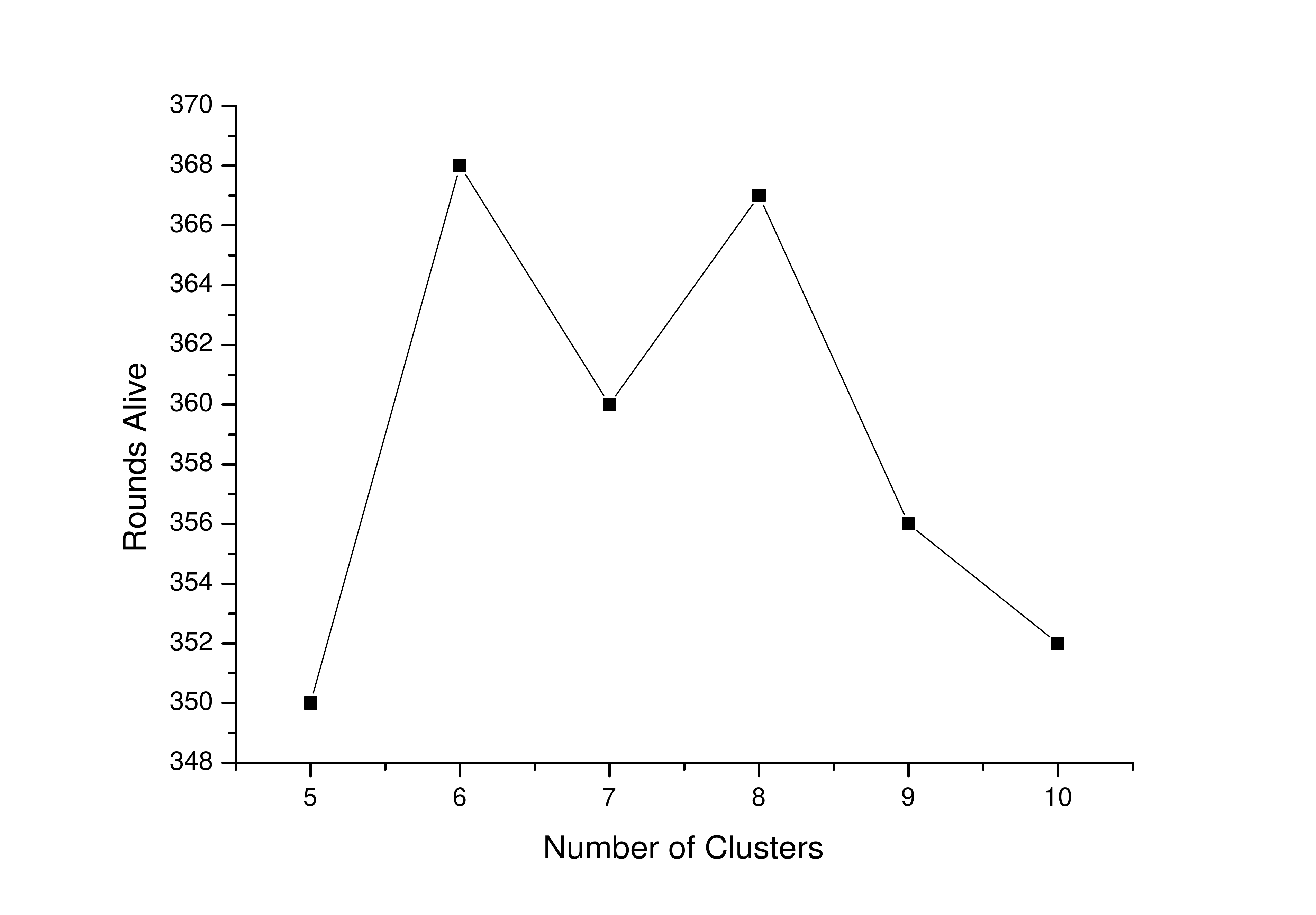}
\end{center}
\caption{Number of rounds alive}
\label{image1b}
\end{figure}

\begin{figure}[tb]
\begin{center}
\includegraphics[width=8cm]{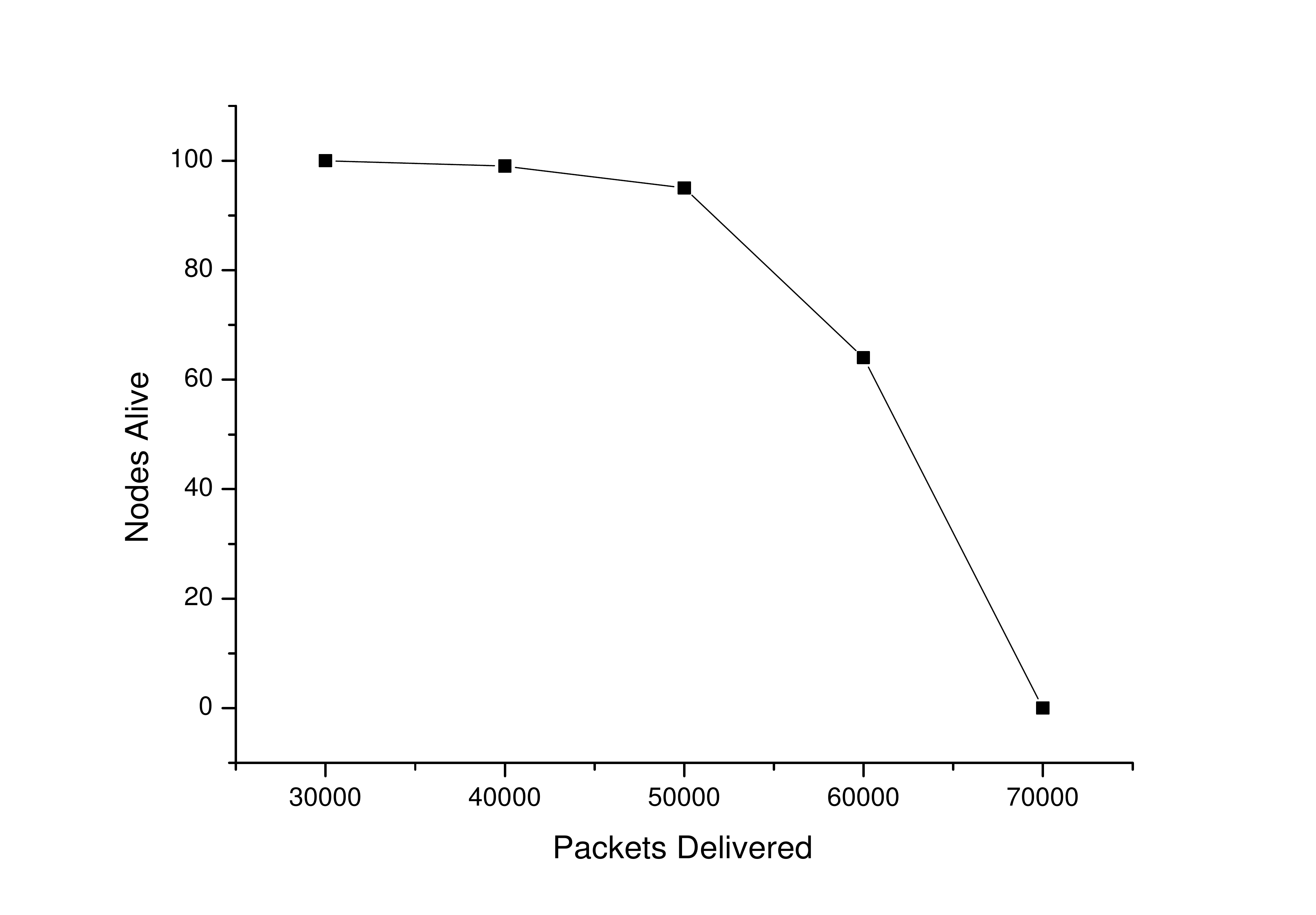}
\end{center}
\caption{Packets delivered}
\label{image1a}
\end{figure}

 Performance is measured by the number of rounds alive and the total data messages successfully delivered. Following figures 1 \& 2 shows the simulated results. It is obvious that our algorithm outperform EECR in the number of rounds the nodes alive. The nodes that remain alive in EECR are a maximum of 175 rounds, whereas with our proposed method rounds alive are 350 ({\it see} Fig. \ref{image1b}). If the system life time is defined as the number of rounds alive, with our proposed technique system life can increase 90\%. Subsequently the number of packets delivered at the base station during the number of rounds of activity is increased from a maximum of 40000 (EECR) to 70000 ({\it see} Fig. \ref{image1a}).
\subsection{Fault Tolerant system}
To design the system as fault tolerant, here the lists of cluster heads on their priority level are being provided; which are obtained from the next representative trajectories. This list will keep by the base station, in any case of fault occurs with the initial cluster heads, as a choice it can go for other alternatives. Following Table 2 shows the list of cluster heads, the number of nodes handled by cluster head and their expected life time. 
\begin{table*}[t1] %
\caption{List of Cluster heads based on Priority}
\begin{center}
\begin{tabular*} {0.83\textwidth}{@{\extracolsep{\fill}}c l l c} \hline
\textbf{Priority} & \textbf{List of Cluster Heads} & \textbf{No. of nodes handles by each cluster head} & \textbf{Expected Life time (in terms of rounds)}
\\\hline 
1 & 0, 34, 69, 39, 33, 68 & 28, 11, 30, 15, 7, 9 & 368\\
2 & {0, 34, 69, 39, 33, 68}	& {36, 6, 27, 15, 7, 9} & 368\\
3 & {1, 68, 0, 34} & {51, 14, 16, 9} & 351\\
4 & {0, 87, 34, 68}	& {60, 6, 11, 23}	& 351\\\hline
\end{tabular*}
\end{center}
\end{table*}

\section{Conclusions}

In this paper, a novel Fault Tolerant Trajectory based clustering solution is presented for selecting cluster heads in wireless sensor networks. Trajectory clustering algorithm enables sensor nodes to reduce data packets by data aggregation. The wireless communication cost is decreased by reduction of data packets, and thus the clustering technique extends the lifetime by reducing the energy consumption of the network. The simulation results demonstrated that our proposal significantly improves the lifetime and reduce the energy consumption of wireless sensor networks compared with existing clustering protocols. We assume that the nodes are error free. However, error will arise due to the noise in the real network environments. As a future work, we plan to extend the method to increase its robustness.

\section*{Acknowledgments}

Hazarath Munaga, would like to thank Prof. Garimella Rama Murthy, Associate Professor, Purdue University, U.S.A for his suggestions during initial stage of this work.


\begin{thebibliography}{99}
\bibitem{1} R. Shah, J. Rabaey. Energy aware routing for low energy ad hoc sensor networks, in: Proc. IEEE Wireless Communications and Networking Conference(WCNC), 2002
\bibitem{2} Y. Yu, R. Govindan, D. Estrin. Geographical and energy aware routing: a recursive data dissemination protocol for wireless sensor networks, 2001
\bibitem{3} J. Chang, L. Tassiulas. Energy conserving routing in wireless ad-hoc networks, in: IEEE Infocom, pp.22--31, 2000
\bibitem{4} J.-H. Chang, L. Tassiulas. Maximum lifetime routing in wireless sensor networks, IEEE/ACM Transactions on Networking, 12(4), pp. 609--619, 2004.
\bibitem{5} F. Ye, G. Zhong, S. Lu, L. Zhang. PEAS: A robust energy conserving protocol for long-lived sensor networks, in: 3rd International Conference on Distributed Computing Systems (ICDCS '03), 2003
\bibitem{6} C. Gui, P. Mohapatra. Power conservation and quality of surveillance in target tracking sensor networks, in: Proc. of the 10th annual international conference on Mobile computing and networking (Mobi-Com '04), ACM Press, New York, NY, USA, pp. 129--143, 2004
\bibitem{7} J. Deng, Y.S. Han, W.B. Heinzelman, P.K. Varshney. Balanced energy sleep scheduling scheme for high density cluster-based sensor networks, Elsevier Computer Communications Journal, Special Issue on ASWN04 28, pp. 1631--1642, 2005 
\bibitem{8} Heinzelman W, Chandrakasan A, Balakrishnan H. Energy efficient communication protocol for wireless microsensor networks. In Proc. of the 33rd Annual Hawaill International Conference on System Sciences, Jan 4--7, 2000, Maui, HI, USA. Los Alamitos, CA, USA: IEEE Computer Society, p: 223, 2000 
\bibitem{9} W. Heinzelman, Application-specific protocol architectures for wireless networks, Ph.D. thesis, Massachusetts Institute of Technology. 
\bibitem{10} S. Lindsey, C. Raghavendra, K. Sivalingam, Data gathering algorithms in sensor networks using energy metrics. IEEE Transactions on Parallel and Distributed System, 13(9), pp. 924--935, 2002
\bibitem{11} O. Younis, S. Fahmy, Heed: A hybrid, energy-efficient, distributed clustering approach for ad hoc sensor networks, IEEE Transactions on Mobile Computing, 3(4), pp. 366--379, 2004
\bibitem{12} C. Li, M. Ye, G. Chen, J. Wu, An energy-efficient unequal clustering mechanism for wireless sensor networks, in: Proc. of the 2nd IEEE International Conference on Mobile Ad-hoc and Sensor Systems (MASS'05), 2005.
\bibitem{13} M. Demirbas, A. Arora, V. Mittal, Floc: A fast local clustering service for wireless sensor networks, in: Workshop on Dependability Issues in Wireless Ad Hoc Networks and Sensor Networks (DIWANS/DSN), 2004.
\bibitem{14} S. Bandyopadhyay, E.J. Coyle, An energy-efficient hierarchical clustering algorithm for wireless sensor networks, in: IEEE INFOCOM, 3pp. 1713--1723, 2003
\bibitem{16} C.-Y. Wen, W.A. Sethares, Automatic decentralized clustering for wireless sensor networks, EURASIP J. Wirel. Commun. Netw., 5(5), pp. 686--697, 2005
\bibitem{17} C. Intanagonwiwat, R. Govindan, D. Estrin, Directed diffusion: a scalable and robust communication paradigm for sensor networks, in: Mobile Computing and Networking, pp. 56--67, 2000
\bibitem{18} B. Krishnamachari, D. Estrin, S.B. Wicker, The impact of data aggregation in wireless sensor networks, in: Proc. of the 22nd International Conference on Distributed Computing Systems (ICDCSW '02), pp. 575--578, 2002
\bibitem{19} T. Kwon and M. Gerla, Clustering with power control, Proc. of IEEE Military Communications Conference (MILCOM), vol. 2, Atlantic City, NJ, pp. 1424--1428, 1999
\bibitem{20} M. Perillo, Z. Cheng, and W. Heinzelman, On the problem of unbalanced load distribution in wireless sensor networks, In IEEE GLOBECOM Wireless Ad Hoc and Sensor Networks, 2004.
\bibitem{21} S. Ahn and D. Kim, Proactive context-aware sensor networks. In European Workshop on Wireless Sensor Networks, Zurich, Switzerland, 2006
\bibitem{22} D. Estrin, R. Govindan, J. S. Heidemann, and S. Kumar. Next century challenges: Scalable coordination in sensor networks. In Mobile Computing and Networking (Mobicom'99), pp. 263--270, 1999
\bibitem{23} N. Bulusu, D. Estrin, L. Girod, and J. Heidemann. Scalable coordination for wireless sensor networks: Self-configuring localization systems. In International Symposium on Communication Theory and Applications (ISCTA), 2001
\bibitem{24} T. Moscibroda, R. Wattenhofer, Maximizing the lifetime of dominating sets, in: 19th International Parallel and Distributed Processing Symposium, 2005
\bibitem{25} S.V. Pemmaraju, I.A. Pirwani, Energy conservation in wireless sensor networks via domatic partitions, in: MOBIHOC, 2006
\bibitem{26} C. Efthymiou, S. Nikoletseas, J. Rolim, Energy balanced data propagation in wireless sensor networks, in: Best papers of the Fourth International Workshop on Algorithms for Wireless, Mobile, Ad Hoc and Sensor Networks, 2004
\bibitem{27} S. Muruganathan, Ma DCF, PI. Bhasin, A centralized energy-efficient routing protocol for wireless sensor networks, IEEE Communications Magazine, 43(3), pp. 8--3, 2005
\bibitem{28} H.O. Tan, I. Korpeoglu, Power efficient data gathering and aggregation in wireless sensor networks, Issue on Sensor Network Technology, SIGMOD Record.
\bibitem{29} Y. Yang, H.-H. Wu, H.-H. Chen, Short: Shortest hop routing tree for wireless sensor networks, in: Proc. of IEEE ICC 2006, 2006
\bibitem{30} LI Li, SONG Shu-Song, Wen Xiang-ming. An energy efficient clustering routing algorithm for wireless sensor networks, The Journal of China Universities of Posts and Telecommunications,13(3), 2006
\bibitem{31} Hazarath Munaga, Lucio Ieronutti, Luca Chittaro, CAST -– A Novel Trajectory Clustering and Visualization tool for Spatio Temporal Data, in the First International conference on Intelligent Human Computer Interaction (IHCI-2009), Springer, pp. 169--175, 2009  
\bibitem{32} Hazarath Munaga, J. V. R. Murthy and N. B. Venkateswarlu, A Novel Trajectory clustering technique for selecting cluster heads in Wireless sensor networks, in the International Journal on Recent Trends in Engineering,1(1), pp. 357--361, 2009
\bibitem{33} E. W. Dijkstra. A note on two problems in connexion with graphs. Numer. Math. (1), pp.269--271, 1959
\bibitem{34} R. Agrawal, C. Faloutsos, and A. N. Swami. Efficient Similarity Search In Sequence Databases, in FODO: Proc. of the 4th International Conference of Foundations of Data Organization and Algorithms, D. Lomet, Ed. Chicago, Illinois: Springer Verlag, pp.69--84, 1993
\bibitem{35} C. Faloutsos, M. Ranganathan, and Y. Manolopoulos. Fast subsequence matching in time-series databases, in SIGMOD '94: Proc. of the 1994 ACM SIGMOD international conference on Management of data. New York, NY, USA: ACM, pp. 419--429, 1994
\bibitem{36} Z. R. Struzik and A. Siebes. Measuring time series' similarity through large singular features revealed with wavelet transformation, in DEXA '99: Proc. of the 10th International Workshop on Database and Expert Systems Applications. Washington, DC, USA: IEEE Computer Society, p. 162, 1999
\bibitem{37} M. Gavrilov, D. Anguelov, P. Indyk, and R. Motwani. Mining the stock market (extended abstract): which measure is best? in KDD '00: Proc. of the sixth ACM SIGKDD international conference on Knowledge discovery and data mining. New York, NY, USA: ACM, pp. 487--496, 2000
\bibitem{38} H. Sakoe and S. Chiba, Dynamic programming algorithm optimization for spoken word recognition. San Francisco, CA, USA: Morgan Kaufmann Publishers Inc., 1990
\bibitem{39} B. Bollobas, G. Das, D. Gunopulos, and H. Mannila. Time-series similarity problems and well-separated geometric sets, in Symposium on Computational Geometry, pp. 454--456, 1997
\bibitem{40} D. Berndt and J. Clifford. Using dynamic time warping to find patterns in time series, AAAI94 Workshop on KDD, 1994
\bibitem{41} D. P. Huttenlocher and K. Kedem. Computing the minimum Hausdorff distance for point sets under translation, in SCG '90: Proc. of the sixth annual symposium on Computational geometry. New York, NY, USA: ACM, pp. 340--349, 1990


\end{thebibliography}
\end{document}